\documentclass[twocolumn,aps,prb,preprintnumbers]{revtex4}
\usepackage{graphicx}       
\usepackage{dcolumn}        
\usepackage{bm}             

\begin{document}


\title{
Tunnel magnetoresistance and interfacial electronic state }

\author{
J. Inoue$^{ac}$ and H. Itoh$^b$}

\affiliation{
$^a$Department of Applied Physics, 
    Nagoya University, Nagoya 464-8603, Japan \\
$^b$Department of Quantum Engineering, 
    Nagoya University, Nagoya 464-8603, Japan \\
$^c$CREST, Japan Science and Technology Corporation (JST)
}

\date{\today}

\begin{abstract}
We study the relation between tunnel magnetoresistance (TMR)
and interfacial electronic states modified by magnetic impurities 
introduced at the interface of the ferromagnetic tunnel junctions, 
by making use of the periodic Anderson model and the linear response theory. 
It is indicated that the TMR ratio is strongly reduced depending
on the position of the $d$-levels of impurities, 
based on reduction in the spin-dependent $s$-electron tunneling 
in the majority spin state. 
The results are compared with experimental results 
for Cr-dusted ferromagnetic tunnel junctions, and 
also with results for metallic multilayers 
for which similar reduction in giant magnetoresistance 
has been reported. 
\end{abstract}

\draft{ICMFS 2002, Kyoto}

\maketitle

\section{Introduction}

The tunnel magnetoresistance (TMR) of ferromagnetic tunnel junctions (FTJs) 
has attracted much interest due to its potentials 
for technological applications.
\cite{miyazaki,moodera1} 
To realize the high MR ratio and low resistance 
required for technological applications, 
more information on the electronic states and tunneling process 
in FTJs is needed. 
Thus far, a simple Julliere's model\cite{julliere,maekawa} 
for the TMR ratio, given as 
\begin{equation}
MR~ratio = \frac{2P_LP_R}{1+P_LP_R},
\end{equation}
where $P_{L(R)}$ is the spin polarization of the left(right)-hand side 
ferromagnetic leads, has frequently been used to analyze the TMR ratio. 
Although this expression well accounts for 
the experimental results of the TMR ratio 
once we adopt the values of the spin polarization $P$ observed,
the meaning of $P$ is not sufficiently clear. 
In order to clarify the relation between the realistic 
electronic state and the TMR ratio, 
first principles and realistic tight-binding  
calculations have been performed to give higher TMR ratios 
than those observed. 
\cite{butler,mathon}
Also, as a result of the roughness of the amorphous-like barrier, 
the effects of resonant tunneling, 
\cite{bratkovsky} 
scattering by disorder, and spin-flip 
tunneling due to magnetic impurities have been intensively studied. 
\cite{guinea,inoue1,itoh1,itoh2}

In spite of these intensive theoretical studies, 
the role of the interface is not yet completely understood.
Recent experiments, however, have indicated the importance of 
the interfacial electronic states: 
Moodera $et~al.$ 
\cite{moodera2}
showed that a formation of quantum well states 
at the interface by Cu spacer decreased the TMR ratio in a 
oscillatory way. 
Teresa $et~al.$ 
\cite{teresa}
showed that the sign of $P$ depends on the combination of the atomic 
species of the ferromagnetic leads and the barrier materials. 
LeClair $et~al.$ 
\cite{leclair}
demonstrated that the Cr-dusted interface 
strongly reduced the TMR ratio of Co/Al-O/Co FTJs, 
whereas the Cu-dusted interface showed a much smaller effect, 
suggesting a close relationship between the electronic density of states 
at interfaces and the TMR ratio of FTJs. 
As for the decay of the TMR ratio due to the quantum well state, 
Zhang and Levy
\cite{levy} proposed that a loss of the coherence due to 
non-ideal structure of the inserted spacer may give rise to an exponentical 
decay of the TMR ratio as a function of spacer thickness.
The difference between Cr and Cu spacers, however, appears even for 
0.1 nm thickness of the spacer, which may indicate that 
a disorderd configuration of atoms at the interface 
must be properly taken into consideration. 

In this paper, we focus our attention on electronic states 
of the magnetic impurities at the interfaces, 
which have partly been argued by LeClair $et~al.$,
\cite{leclair}
and will show that the observed phenomena 
for Cr-dusted FTJs can be explained 
by the concept that the $s$-electron tunneling, which is dominant 
in the tunneling process, 
is strongly affected by changes induced by magnetic impurities 
in the electronic states at the interfaces. 
Below we adopt the periodic Anderson Hamiltonian to model 
the impurity-doped Co/Al-O/Co type FTJs, 
calculate the tunnel conductance and TMR ratio by using the Kubo formula, 
and show that the TMR ratio of FTJs made of Co or permalloy is 
actually sensitive to the majority spin electronic states induced 
by magnetic impurities at interfaces. 
Strong reduction in the giant magnetoresistance (GMR) or inverse GMR induced 
by magnetic impurities in metallic multilayers has already been reported.
\cite{hsu,vouille,marrows} 
Such reduction in GMR ratios has been interpreted in terms of the change 
in the spin asymmetry parameter.
\cite{campbell} 
We calculate the effects of magnetic impurities on the GMR ratio in 
a generalized periodic Anderson model ($s-d$ model), 
then compare the effects of magnetic impurities between TMR and GMR . 

\section{Model and formalism}

In order to describe TMR and the electronic states of 
magnetic impurities in the interfacial Co layers in Co/Al-O/Co type FTJs,
we must take into account the following two aspects. 
First, the positive spin polarization of the tunnel current 
and the reasonably large TMR ratio of Co/Al-O/Co type FTJs 
must be properly reproduced
in the model. 
The simplest way to satisfy these conditions is to use $s$-electron 
tunneling that is spin-polarized by spin-dependent $s-d$ mixing.
Second, it is well known that transition metal impurities 
such as Cr and V in Co metal have magnetic moments 
antiparallel to the Co moment, 
and form the so-called virtual bound state (VBS) in the majority spin 
state of Co.
\cite{friedel,hasegawa}
Because the formation of VBS results from the mixing 
of the impurity $d$-level with the $s$-band of Co metal, 
we inevitably include the $s$-band in the model as well as the $d$-band 
in order to describe the ferromagnetism of Co leads. 
Thus, the periodic Anderson model can include the basic features 
necessary to elucidate the electronic states 
and the characteristics of TMR. 

The Hamiltonian of the periodic Anderson model is given by 
\begin{equation}
H=H_s + H_d + H_{sd}.
\end{equation}
$H_s$ is the Hamiltonian for conduction electrons in a tight-binding 
version given as 
\begin{equation}
H_s = -t^s \sum_{ij\sigma} c_{i\sigma}^{\dagger}c_{j\sigma} 
+ \sum_{i\sigma}v^sc_{i\sigma}^{\dagger}c_{i\sigma},
\end{equation}
where $t^s$ is the nearest neighbor (n.n.) hopping integral 
for $s$-electrons being independent of materials, 
$v^s$ is the energy level of the $s$-electrons, 
and $\sigma $ indicates the spin ($\uparrow$ or $\downarrow$).
The second term of eq.(2) indicates the $d$-levels, and is given as
\begin{equation}
H_d = \sum_{i\sigma}v_{i\sigma}^dd_{i\sigma}^{\dagger}d_{i\sigma}.
\end{equation}
The last term indicates the $s-d$ mixing, as 
\begin{equation}
H_{sd} = - \gamma \sum_{i\sigma}(c_{i\sigma}^{\dagger}d_{i\sigma} + h.c.).
\end{equation}
We neglect $d-d$ hopping because the tunneling is dominated 
by $s-s$ hopping, 
and assume that, by taking $\gamma =0$ in the barrier region, 
only $s$-electrons tunnel through the barrier.
Because we deal with transition metal impurities, $v_{i\sigma}^d$ 
represents both the $d$-levels $v_{\sigma}^d$ of the 
ferromagnetic leads and those of 
the impurities $V_{{\rm imp}\sigma}$ introduced at the interface. 

The tunnel conductance at the zero-bias limit 
is calculated in numerical simulations using the Kubo formula.
\cite{itoh1}
We adopt a simple cubic structure for a finite-sized system; 
the cross section of the system includes $12 \times 12$ sites 
and the barrier thickness 
is 4 in units of the lattice constant $a$ .
Thereafter, the parameter values are taken 
in units of $t^s$, such that $v^s=0.0$, 
$v_{\uparrow}^d=-2.5$, $v_{\downarrow}^d=-2.0$,  
and $\gamma =1.5$ for ferromagnetic leads. 
The barrier potential is taken to be 9.0.
As can be seen later, a rather large value of $\gamma$ is required 
to obtain an appreciable TMR ratio by the $s-d$ mixing. 

\section{Calculated results}

We first calculate the tunnel conductance without impurities, 
$\Gamma_{\rm P}$ and $\Gamma_{\rm AP}$, 
in parallel (P) and antiparallel (AP)
alignment of the magnetizations of the ferromagnetic leads, respectively.
The TMR ratio is defined as 
\begin{equation}
MR~ratio = \frac{\Gamma_{\rm P}-\Gamma_{\rm AP}}{\Gamma_{\rm P}} \,.
\end{equation}
Figure 1 shows the calculated results of tunnel conductance as a function of 
the Fermi energy $\epsilon_{\rm F}$. 
\begin{figure}[tb]
\begin{center}
\leavevmode
\includegraphics[width=0.95\linewidth,clip]{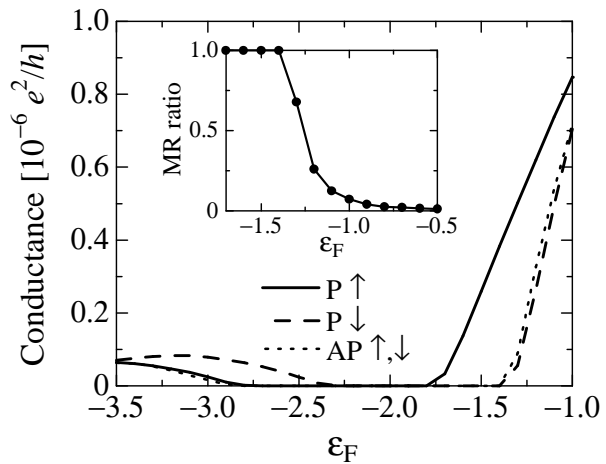}
\caption{
Calculated results of tunnel conductance as a function of the Fermi level 
$\epsilon_{\rm F}$ for parallel (P) and antiparallel (AP) alignment 
of the magnetization of the ferromagnetic leads. 
Inset: TMR ratio calculated from the tunnel conductance 
as a function of $\epsilon_{\rm F}$.}
\label{fig1}
\end{center}
\end{figure}
As a result of the hybridization gap, 
$\Gamma_{\rm P}$ is zero between $\epsilon_{\rm F}=-2.75$ and $-1.80$ 
for $\uparrow$ spin channel, 
and between $\epsilon_{\rm F}=-2.25$ and $-1.40$ 
for $\downarrow$ spin channel. 
Consequently, $\Gamma_{\rm AP}=0.0$ for $-2.75<\epsilon_{\rm F}<-1.40$.
The inset of the figure shows the results of the TMR ratio 
as a function of $\epsilon_{\rm F}$.
The high TMR ratio below $\epsilon_{\rm F}=-1.4$ is due to opening of the 
hybridization gap, but there is no current in AP alignment 
in this energy region. 
Above $\epsilon_{\rm F}=-1.0$, there is almost no TMR effect, because 
the current is carried by spin-unpolarized $s$-electrons.
The physically meaningful values of the TMR ratio appear 
in a rather narrow range of 
energy, $(-1.3 < \epsilon_{\rm F} < 1.1)$ where 
the $s$-electrons of the minority spin
states carry less current than the majority $s$-electrons 
due to the stronger $s-d$ mixing at $\epsilon_{\rm F}$. 
Because the range in which a meaningful TMR ratio appears 
becomes narrower with decreasing $\gamma$, 
we take $\epsilon_{\rm F}=-1.2$ with a rather large value of 
$\gamma=1.5$ hereafter, which brings about 
an $MR~ratio$ of $\sim 0.25$, which is not unreasonable 
when compared with the experimental values.

Now we introduce impurities into a ferromagnetic metal-layer adjacent 
to the insulating barrier. 
Because the transition metal impurities of the Cr type 
in which we are interested may form magnetic moments antiparallel to the 
magnetic moments of the ferromagnetic leads made, for example, of Co, 
there is a relation such that $V_{{\rm imp}\uparrow}>V_{{\rm imp}\downarrow}$.
Because of the relation, 
the up spin state of the impurities forms VBS, while the down spin state 
is almost occupied. 
Therefore, in order to realize the situation, 
we take $V_{{\rm imp}\downarrow}=-2.5$, which is slightly 
below $v_{\downarrow}^d=-2.0$,
and treat $V_{{\rm imp}\uparrow}$ as a variable parameter.
In order to gain insight into the electronic state of the impurity, 
we calculated the density of states (DOS) of a single impurity 
on the metallic surface, which value 
is a reasonable approximation of the impurity DOS at the interface of FTJs. 
We show in the inset of Fig. 2 the dependence of the number of $\uparrow$ 
spin electron $n_{{\rm imp}\uparrow}$ of the impurity 
as a function of $V_{{\rm imp}\uparrow}$.
\begin{figure}[tb]
\begin{center}
\leavevmode
\includegraphics[width=0.95\linewidth,clip]{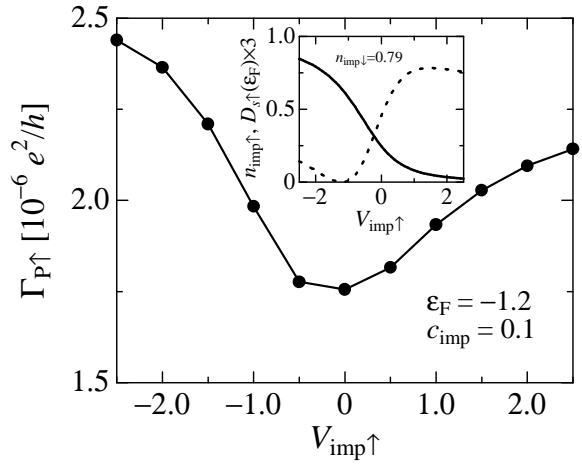}
\caption{
Calculated results of tunnel conductance in the parallel alignment 
as a function of the impurity potential $V_{{\rm imp}\uparrow}$ 
for $\epsilon_{\rm F}=-1.2$ with $c_{\rm imp}=0.1$. 
Inset: the number of up spin electrons $n_{{\rm imp}\uparrow}$ (solid curve)
and $s$-component of the density of states (broken curve) of a single 
impurity on a ferromagnetic surface as a function of $V_{{\rm imp}\uparrow}$.}
\label{fig2}
\end{center}
\end{figure}
With increasing $V_{{\rm imp}\uparrow}$, $n_{{\rm imp}\uparrow}$ decreases, 
indicating that the spin polarization of the impurity becomes negative for 
larger values of $V_{{\rm imp}\uparrow}$ 
because $n_{{\rm imp}\downarrow}\sim 0.79$.
Because the drop in $n_{{\rm imp}\uparrow}$ becomes stronger 
when $V_{{\rm imp}\uparrow}$ approaches $\epsilon_{\rm F}=-1.2$, 
we find that the impurity DOS or VBS is located near $\epsilon_{\rm F}$.
Although we did not carry out self-consistent calculations for 
the impurity state, the magnetic moment of the impurity is about 0.4 
per $d$-orbital for $V_{{\rm imp}\uparrow}=-0.5$ and 
$V_{{\rm imp}\downarrow}=-2.5$, which gives an on-site Coulomb interaction 
of $\sim 5$ in units of $t^s$, corresponding to $\sim $1eV 
for 3$d$ transition metals when we take $t^s \sim 1$eV. 
It should be noted, however, that the spin polarization of the 
ferromagnetic leads is rather small; 
this is a shortcoming of the present model, which was adopted to reproduce 
both positive spin polarization and a reasonable TMR ratio.

The tunnel conductance of FTJs including impurities 
with concentration $c_{\rm imp}$ is calculated 
by averaging over ten samples for different impurity-configurations.
The root-mean-square deviation of the conductance is 
almost one order of magnitude smaller than the conductance itself. 
Figure 2 shows the tunnel conductance $\Gamma_{{\rm P}\uparrow}$ 
in the parallel alignment as a function of $V_{{\rm imp}\uparrow}$. 
It can be seen that the conductance decreases pronouncedly 
when $V_{{\rm imp}\uparrow}$ exceeds $\epsilon_{\rm F}$.
This is because the $s-d$ mixing decreases the $s$-component 
of the majority spin state near $\epsilon_{\rm F}$
as the VBS approaches $\epsilon_{\rm F}$, 
and blocks the $s$-electron tunneling. 
In other words, an anti-resonance of the $s$-states appears
at the interfacial layer where the impurities are introduced.
The broken curve in the inset of Fig. 2 shows the $s$-component of 
the DOS of an impurity on a surface of an electrode. 
It can be seen the anti-resonant state appears 
when $V_{{\rm imp}\uparrow} \sim \epsilon_{\rm F}$,
although the correspondence between the $V_{{\rm imp}\uparrow}$-dependences 
of the DOS and $\Gamma_{\rm P}$ is not perfect, probably due to 
the opening of the hybridization gap in the periodic Anderson model.

As a consequence of the reduction in the $\uparrow$ spin 
conductance in P alignment, the TMR ratio dramatically reduces as 
$V_{{\rm imp}\uparrow}$ approaches $\epsilon_{\rm F}$, 
as shown in Fig. 3, where the results for $c_{\rm imp}=0.2$ 
as well as those for $c_{\rm imp}=0.1$ are presented.
\begin{figure}[tb]
\begin{center}
\leavevmode
\includegraphics[width=0.95\linewidth,clip]{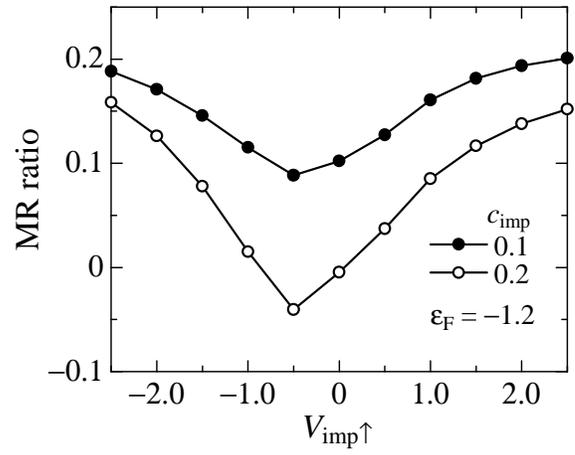}
\caption{
TMR ratio as a function of $V_{{\rm imp}\uparrow}$ for 
$\epsilon_{\rm F}=-1.2$.
Solid and open circles are the results for $c_{\rm imp}=0.1$ 
and $0.2$, respectively.}
\label{fig3}
\end{center}
\end{figure}
The results indicate that the TMR ratio decreases when 
VBS appears near the Fermi level and weakens the $s$-electron tunneling.   
The reduction of the TMR ratio due to magnetic impurities, 
however, is quantitatively stronger than the experimental value. 
The might be due to the rather strong $s-d$ mixing 
chosen in this model. 

The relative position of VBS of magnetic impurities 
in transition metals has been intensively studied before;
\cite{friedel,hasegawa,podloucky}
VBS for Cr impurities is close to the Fermi level, 
whereas that of V impurities is higher than the Fermi level.
For Cu impurities, on the other hand, $V_{{\rm imp}\uparrow}$ 
may be nearly the same with $v_{\uparrow}^d=-2.5$, and does not form 
any VBS. 
Therefore, we may conclude that impurities corresponding to Cr and V 
may bring about dramatic reduction 
in the TMR ratio whereas Cu impurities do not. 
The results of both the conductance and TMR ratio 
are thus in good agreement with the experimental observations.
\cite{leclair}
Further, the present model showed no reduction in 
the conductance of the $\uparrow$ spin state 
when magnetic impurities were placed on the Co leads several 
layers apart from the interface. 
Rather, it slightly increased near $V_{{\rm imp}\uparrow} \sim 0$.
This result is not inconsistent with the experimental result. 
\cite{leclair}

\section{Comparison with CIP-GMR}

In order to compare the results obtained above with 
similar effects reported for GMR, we performed simple calculations 
of GMR in trilayers doped with magnetic impurities near the interface.
As a simple model of Co/Cu/Co type trilayers, 
we adopted the $s-d$ model, 
the Hamiltonian of which is given by adding 
the following Hamiltonian to eq. (2):
\begin{equation}
H_d = -t^d\sum_{ij\sigma}d_{i\sigma}^{\dagger}d_{j\sigma},
\end{equation}
because $d$-electrons may carry the current in this case.
The n.n. hopping integral for $d$-electrons is taken to be $0.1$.
Now the $d$-band is broadened by the $d$-electron hopping, and we 
take a smaller value of the $s-d$ mixing; $\gamma =0.3$. 

The lattice structure is assumed to be simple cubic, 
and the thickness of each layer is /6/12/6/ atomic layers 
in units of the lattice constant.
Instead of the site representation used for TMR case, 
we use a mixed representation $(l,{\bf k}_{\parallel})$ to treat 
the semi-infinite systems where $l$ is an index of the atomic layers 
and ${\bf k}_{\parallel}=(k_x, k_y)$ is a Fourier transform 
of the sites on each atomic layer. 
The Green's function ${\bf G}_{\sigma}(z)=(z-H_{\sigma})^{-1}$ 
with $z=\epsilon +i\eta$ is expressed by a tridiagonal matrix 
where the diagonal elements consist of $2 \times 2$ matrices of
\begin{equation}
g_{l\sigma}^{-1}({\bf k}_{\parallel}) = \pmatrix{ 
z-\epsilon_{{\bf k}_{\parallel}}^s &
-\gamma \cr
-\gamma  &
z-\epsilon_{{\bf k}_{\parallel}}^d - v_{l\sigma}^d \cr
 }, 
\end{equation}
and the off-diagonal elements are alos given by $2 \times 2$ matrices 
including the hopping integrals of $s$- and 
$d$-electrons. 
The $d$-level $v_{i\sigma}^d$ depends on 
the layer, and is expressed as $v_{l\sigma}^d$ in eq. (8).
As in the TMR case, we take $v_{l\uparrow}^d=-2.5$ 
and $v_{l\downarrow}^d=-2.0$ 
when $l$ belongs to the ferromagnetic (Co) layers, 
and $v_{l\uparrow}^d=v_{l\downarrow}^d=-2.5$ when 
$l$ belongs to the nonmagnetic (Cu) layers. 
Because a set of parameter values different from those for the TMR case 
is used for the GMR case, 
the position of $\epsilon_{\rm F}$ is different 
from that taken for the TMR calculation.

To reproduce the GMR effect in our simple model, 
we introduce the interfacial spin-dependent scattering
caused by an intermixing layer at interfaces.
\cite{inoue2,inoue3,levy,gijs,tsymbal} 
Because the $d$-level of 
the minority spin state of Co atoms is above the Cu $d$-band, 
Co atoms intermixed with a Cu layer will form 
VBS in the minority spin state of the Cu $s$-band,
resulting in spin-dependent scattering to produce the GMR. 
Here, we introduce the following self-energy into the $s$-component 
of the Green's function:
\begin{equation}
\Sigma_{\sigma}(z)=\frac{c\gamma^2}{z-v_{l\sigma}^d+i\Delta}, 
\end{equation}
which represents the effects of Co atoms dissolved into the Cu layer. 
Here $c$ indicates the concentration of such Co atoms 
and $\Delta$ is the broadening of $d$-levels due to the $s-d$ mixing, 
which is taken to be $\Delta =0.1$, keeping in mind the width of 
VBS calculated in the first principles.
\cite{podloucky} 
We further take $\eta = 0.001$ to reproduce the spin-independent resistivity. 
Here, we note that the concentration $c$ itself does not have 
sufficient physical meaning, but $c\gamma^2$ controls the magnitude 
of the self-energy. 
The conductivity parallel to the layers can be calculated 
using the Kubo formula 
without vertex corrections, because there is translational invariance 
along the planes. 
The GMR ratios thus calculated increase with increasing $c$ and 
reach about 60\% for $c=0.5$.

After successfully reproducing the GMR effect in our simple model, 
we introduce additional magnetic impurities 
into the adjacent layers of the intermixed interface. 
For simplicity, we use the same expression as that in eq. (9) 
for the self-energy for the majority-spin state of the $s$-band 
caused by the additional magnetic impurities, 
while $c$ and $v_{l\sigma}$ are replaced with 
$c_{\rm imp}$ and $V_{{\rm imp}\sigma}$, respectively. 
As in the calculation of TMR, we take 
$V_{{\rm imp}\downarrow}=v_{l\downarrow}^d$ 
and treat $V_{{\rm imp}\uparrow}$ as a variable parameter. 
The GMR ratios calculated for $c=c_{\rm imp}=0.5$ 
with $\epsilon_{\rm F} =-2.0$ 
are shown in Fig. 4 (open circles) 
as a function of the impurity potential $V_{{\rm imp}\uparrow}$. 
\begin{figure}[tb]
\begin{center}
\leavevmode
\includegraphics[width=0.95\linewidth,clip]{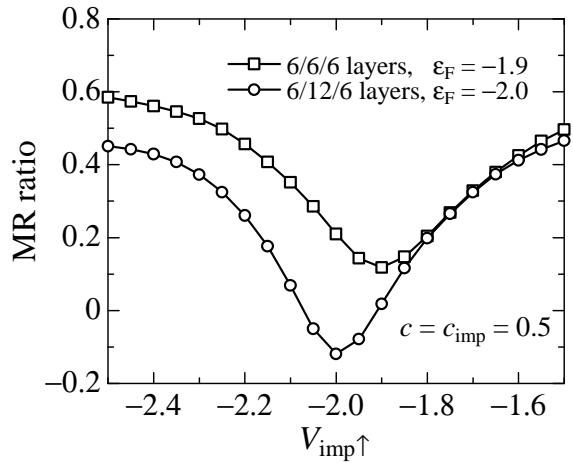}
\caption{
Calculated results of the GMR ratio as a function of the impurity potential 
$V_{{\rm imp}\uparrow}$ for 6/12/6/ and 6/6/6 trilayers, respectively. }
\label{fig4}
\end{center}
\end{figure}
It can be seen that an inverse GMR can be realized 
when $V_{{\rm imp}\uparrow} \sim \epsilon_{\rm F} $. 
Open squares indicate the results for a trilayer with 6/6/6 atomic layers
with $\epsilon_{\rm F} =-1.9$, and the results show that 
the qualitative features are not affected by the thickness of the trilayer.
The GMR ratio becomes minimal when 
$V_{{\rm imp}\uparrow} \sim \epsilon_{\rm F}$, 
which characteristic is also found in TMR systems. 
The reduction, however, may be brought about by an increased scattering of 
$s$-electrons in the majority spin band due to 
$s-d$ mixing with the impurity states in addition to a reduction in the 
$s$-component of the DOS. 
The effects of magnetic impurities such as Cr and V 
in the GMR case are almost the same as those obtained in the TMR case, 
though the origin of the reduction in TMR ratio is due to 
the blockade of $s$-electron tunneling caused by 
the anti-resonance in the majority spin state of $s$-band. 

\section{Conclusion}

In conclusion, by a qualitative and systematic study of the 
dependence of the MR ratio on the type of impurities at the interface, 
we showed that both TMR and GMR ratios are affected strongly 
when the magnetic impurities form VBS in the majority spin state 
near the Fermi level. 
The results are in good agreement with the observed ones. 
Although the effects of magnetic impurities such as Cr and V 
are quite similar on TMR and GRM, the reduction in TMR ratios is due to 
a blockade of $s$-electron tunneling, 
whereas that in GMR is mainly due to an increase 
of $s$-electron scattering in the majority spin state.
More realistic and self-consistent calculations, however, are desired 
to obtain a quantitative understanding between the MR ratio 
and the interfacial electronic states. 

\begin{acknowledgments}
It is a great pleasure for the authors to dedicate this paper 
to Prof. T. Shnjo on the occasion of his retirement from Kyoto University. 
J. I. is grateful for financial support provided by 
the NEDO International Joint Research Project (NAME).
\end{acknowledgments}


\end{document}